\documentclass[pra,singlecolumn,superscriptaddress]{revtex4}
\usepackage{color}
\usepackage{amsmath,amssymb,graphicx,bm,float}

\begin{document}
\title{A Time-Multiplexed Heralded Single-Photon Source}
\author{Fumihiro Kaneda} 
\affiliation{Department of Physics, University of Illinois at Urbana-Champaign, Urbara, IL 61801, USA}
\author{Bradley G. Christensen} 
\affiliation{Department of Physics, University of Illinois at Urbana-Champaign, Urbara, IL 61801, USA}
\author{Jia Jun Wong} 
\affiliation{Department of Physics, University of Illinois at Urbana-Champaign, Urbara, IL 61801, USA}
\author{Kevin T. McCusker} 
\affiliation{Department of Physics, University of Illinois at Urbana-Champaign, Urbara, IL 61801, USA}
\affiliation{EECS Department, Center for Photonic Communication and Computing, Northwestern University, 2145 Sheridan Road, Evanston, Illinois 60208-3118, USA}
\author{Hee Su Park} 
\affiliation{Korea Research Institute of Standards and Science, Daejeon 305340, South Korea}
\author{Paul G. Kwiat} 
\affiliation{Department of Physics, University of Illinois at Urbana-Champaign, Urbara, IL 61801, USA}
\maketitle

\textbf{Photons have proven to be excellent carriers of quantum information, and play essential roles in numerous quantum information processing (QIP) applications \cite{Gisin:2002gb, OBrien:2007io, Childs:2013hh,Pan:2012kv}. 
In particular, heralded single-photon sources  \cite{Hong:1986va} via spontaneous parametric-down conversion (SPDC) have been a key technology for demonstrating small-scale QIP, yet their low generation efficiency is a critical limitation for further scaling up optical QIP technology. 
In order to efficiently overcome the probabilistic nature of SPDC, here we demonstrate time multiplexing \cite{Pittman:2002dx, Jeffrey:2004ky, McCusker:2009ci, Glebov:2013hr,Mower:2011kp}  for up to 30 time slots of a periodically pumped heralded single-photon source, using a switchable low-loss optical storage cavity. 
We observe a maximum single-photon probability of 38.6$\pm$0.4\% in periodic output time windows, corresponding to $\sim$6 times enhancement over a non-multiplexed source, but with no increase in the contribution of unwanted multi-photon events. 
Combining this time-multiplexing technique with a heralded source producing pure single-photon states \cite{Mosley:2008ki, Evans:2010jn, Spring:2013tv} should enable larger scale optical QIP systems than ever realized. 
}

Realizing and scaling up optical QIP systems requires on-demand preparation of quantum states of light such as single-photon and definite multi-photon states. 
Single atoms, ions, and solid-state single-emitter sources such as color-centers in diamond and semiconductor quantum dots can generate true single-photon states, and especially solid-state systems have great potential for integration. 
However, while high indistinguishability \cite{Sipahigil:2014gh,Wei:2014dg} and somewhat high (50-80\%) collection efficiency \cite{Gazzano:2013cq} have been reported in different single-emitter systems, achieving both simultaneously remains a challenge. 
Moreover, most single-emitter sources generate single photons with narrow bandwidths (5-100 MHz) that may be unsuitable for pursuing high-speed applications (e.g., a 5-MHz bandwidth source can produce non-overlapping single-photon gaussian wavepackets---of duration 90 ns---at a maximum rate of $\sim$10 MHz). 

SPDC is another approach that has been conventionally and widely used for generating entangled photon pairs, multi-photon entangled states, and small-scale quantum algorithms \cite{Pan:2012kv}. 
A photon pair generated via SPDC can also be used for generating a ``heralded'' single-photon state; detecting one of the photons ``heralds'' the presence of the other. 
Moreover, current technology has realized photons with very high coupling efficiency into single-mode optical fibers \cite{Christensen:2013ux,Pereira:2013va} and controlled two-photon spectral shape \cite{Mosley:2008ki, Evans:2010jn}. 
However, the photon-pair generation process via SPDC is probabilistic: one cannot obtain a photon pair or a heralded single photon on demand. 
Unfortunately, indefinitely increasing the mean number of  photon pairs per pump pulse $p$ also increases the likelihood of unwanted $k$-photon pairs ($\sim $$p^k$) by higher-order processes. 


To overcome the probabilistic nature of photon-pair generation, time-multiplexing techniques were first proposed and demonstrated by Pittman, Jacobs, and Franson \cite{Pittman:2002dx} in 2002.  
The method was since extended and theoretically analyzed \cite{Jeffrey:2004ky, McCusker:2009ci, Glebov:2013hr,Mower:2011kp} 
(a related approach uses spatial multiplexing \cite{migdall:2002hk,Ma:2011in,Collins:2013eu, FrancisJones:2014ub, Bonneau:2015ho}, but this is much more resource intensive--- a source analogous to ours would need 30 crystals and detectors). 
The basic idea is shown in Fig. 1. (a). 
A laser pulse train with period $\tau$ pumps a $\chi^{(2)}$ nonlinear crystal, generating photon pairs (i.e., signal and idler photons) in one or more time slots. 
Each signal photon is sent to a single-photon detector (SPD) whose firing heralds in which time slot the corresponding idler photon is present. 
By using a switchable storage cavity with the matched cycle length $\tau$, any one of the time slots heralded to contain an idler photon can be multiplexed onto a single output time window (see Fig. 1 (b)). 
Thus, the multiplexed single-photon probability $P_M(1)$ during the output time window is increased according to the number of pump pulses (time slots) $N$ used for one cycle of the multiplexing. 
Moreover, if $N$ is large, the probability of generating unwanted multiple pairs in a given time slot can be made arbitrarily small, because the total pump energy through the multiplexing cycle is distributed over the $N$ time slots, and the ratio of the single- and multi-photon probability is as low as the one for a single (non-multiplexed) heralded single-photon source. 
Hence, assuming losses can be kept low, the multiplexed heralded single-photon source can work as a pseudo-on-demand single-photon source (i.e., $P_M(1) \rightarrow 1$) by using sufficiently low pump pulse energy ($p < 0.1$) and large cycle number $N$. 
A detailed mathematical calculation of $P_M(1)$, accounting for loss and inefficiencies of SPDs, PC, and the SPDC source is discussed in \cite{Jeffrey:2004ky, McCusker:2012wi} and in the Supplementary Information.


\begin{tiny}
\begin{figure}[t!]
  \includegraphics[width=0.54\columnwidth,clip]{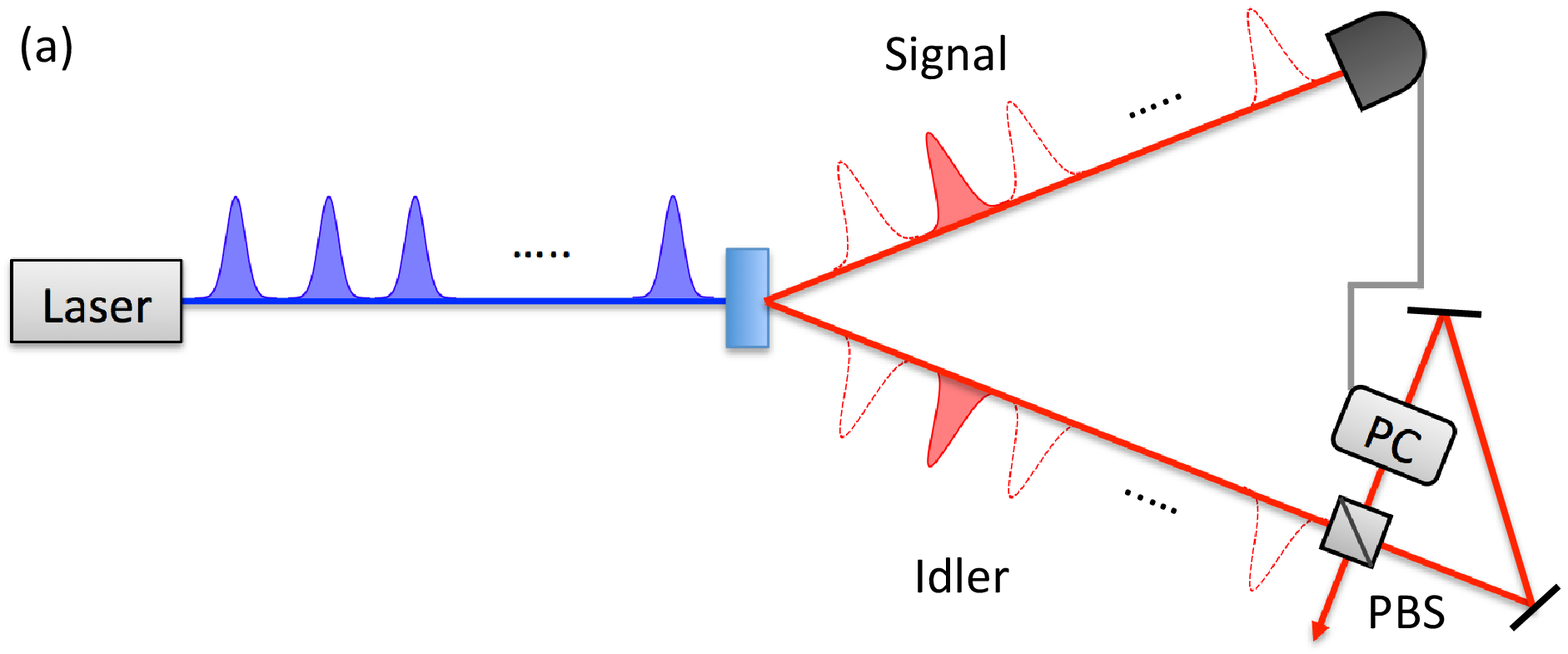}
  \includegraphics[width=0.45\columnwidth,clip]{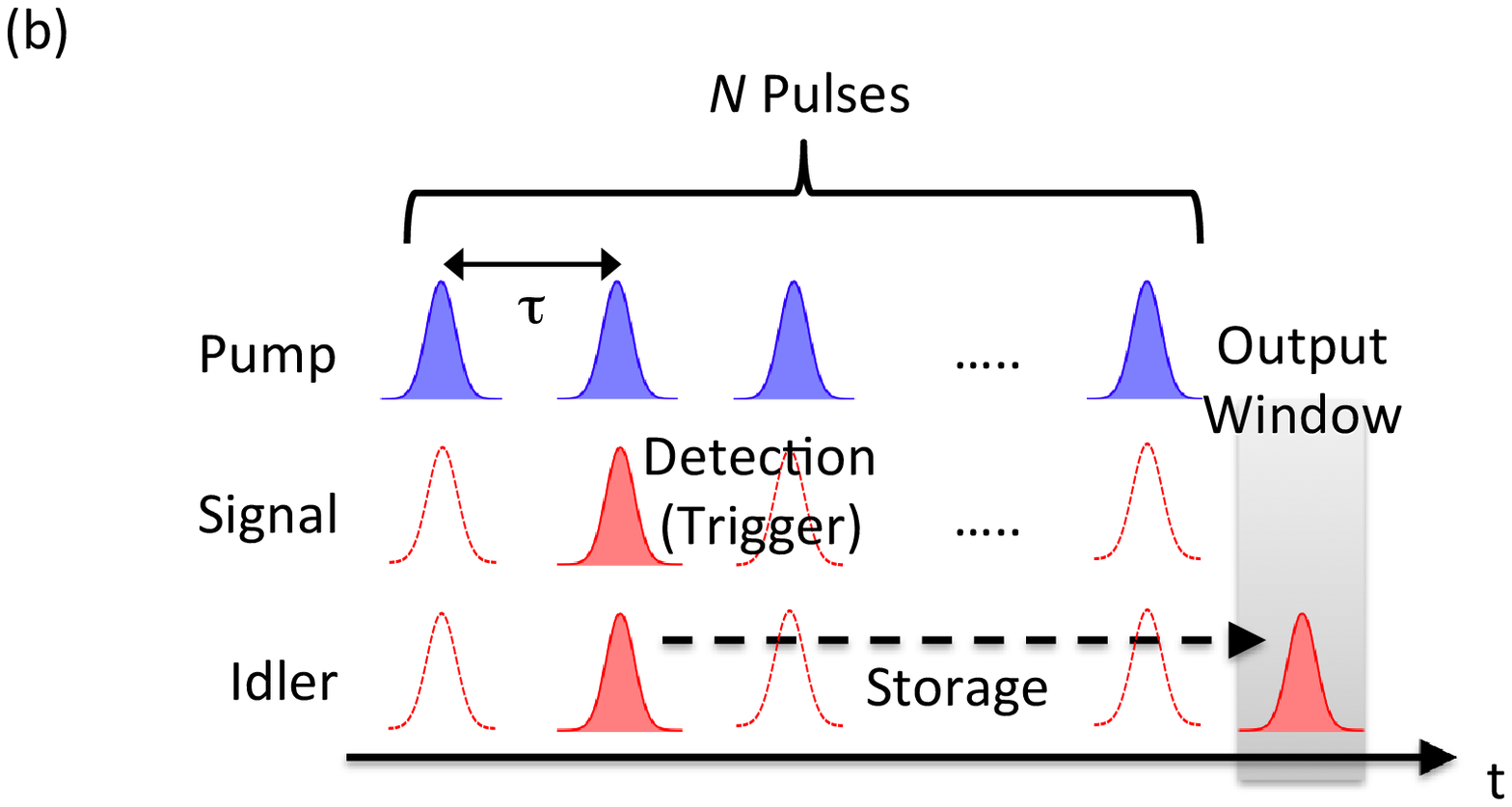}   
   \caption{(a) Simplified schematic diagram of a time multiplexed heralded single-photon source. (b) Timing diagram of pump, signal, and idler photons in the time multiplexing. In our system, when the idler photon enters the storage cavity via a polarizing beam-splitter (PBS), a Pockels cell (PC) in the cavity is fired, rotating the photon polarization by $90\textbf{}^{\circ}$, thereby storing the photon in the cavity loop until its polarization is switched back. 
The single-pass delay of the cavity is set to match $\tau$ so that the time slots where idler photons may be present are temporally and spatially overlapped with each other in the  cavity. 
The idler photon is then released from the cavity in the desired output time slot by a second switching of the PC. 
   }
\label{fig.model}
\end{figure}
\end{tiny}



A schematic diagram of our experimental setup is shown in Fig. 2 (a). 
We implemented the setup with a low-loss heralded single-photon source and photon storage cavity (see Methods).  
However, in practice, many repeated passes through the storage cavity attenuate the stored idler photons; for the case that multiple time slots within $N$ are heralded, photons stored later---closer to the output time window---experience less loss. 
In order to select only the last-produced, and therefore, lowest-loss photon, we introduced a low-loss Herriott-cell-type optical delay line \cite{Robert:2007tj} (between the SPDC crystal and the switchable storage cavity) whose delay $\Delta t_{DL}$ is long enough to contain idler photons from all $N$ of the pump pulses. 
The long optical delay makes it possible to determine the time slot of the last-detected signal photon, and to only switch the corresponding idler photon into the storage cavity. 
For our largest delay, $\Delta t_{DL}$ is $\sim$400 ns, sufficiently longer than the  $\sim$120-ns electronics latency (from detecting signal photons to the PC driver) to hold idler photons from up to $N\sim$30 ($30\times \tau = 250$ ns for $\tau = 8.33$ ns in our experiment). 
Using flippable mirrors, we can also reduce $\Delta t_{DL}$ to $\sim$200 ns, which is still longer than the electronics latency, but not enough to store the latest heralded time slot for larger $N$. 
This allows us to compare the effect of storing the photons in the latest and earliest heralded time slots. 
%

%

\begin{tiny}
\begin{figure}[t!]
  \includegraphics[width=0.8\columnwidth,clip]{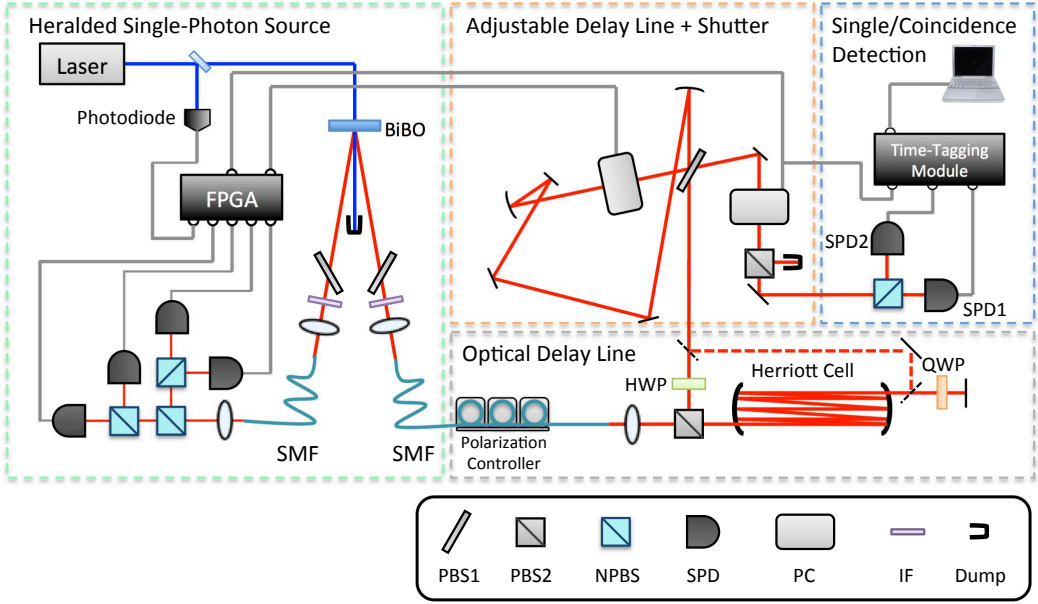}
   \caption{Schematic diagram of experimental setup. PBS1: Brewster-angled polarizing beam splitter. PBS2: cube polarizing beam splitter. NPBS: non-polarizing beam splitter. SPD: single-photon detector. PC: Pockels cell. IF: interference filter ($\Delta \lambda =  20$ nm). HWP: half-wave plate. QWP: quater-wave plate. FPGA: field-programmable gate array. Dashed red line shows the optical path when the optical delay is shortened to $\Delta t_{DL}\sim$200 ns. }
\label{fig.model}
\end{figure}
\end{tiny}

%
\begin{tiny}
\begin{figure}[t!]
  \includegraphics[width=0.31\columnwidth,clip]{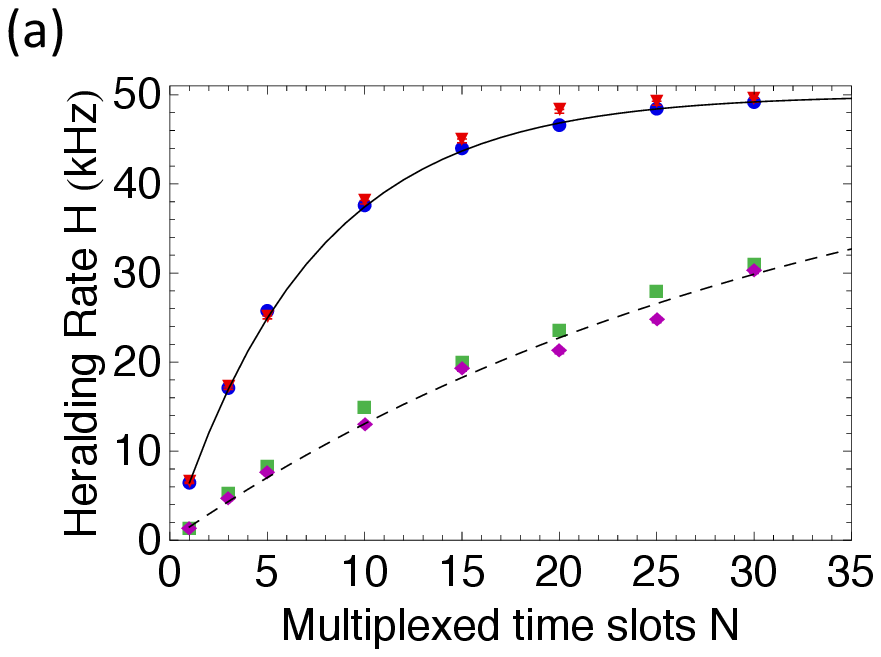}
  \includegraphics[width=0.32\columnwidth,clip]{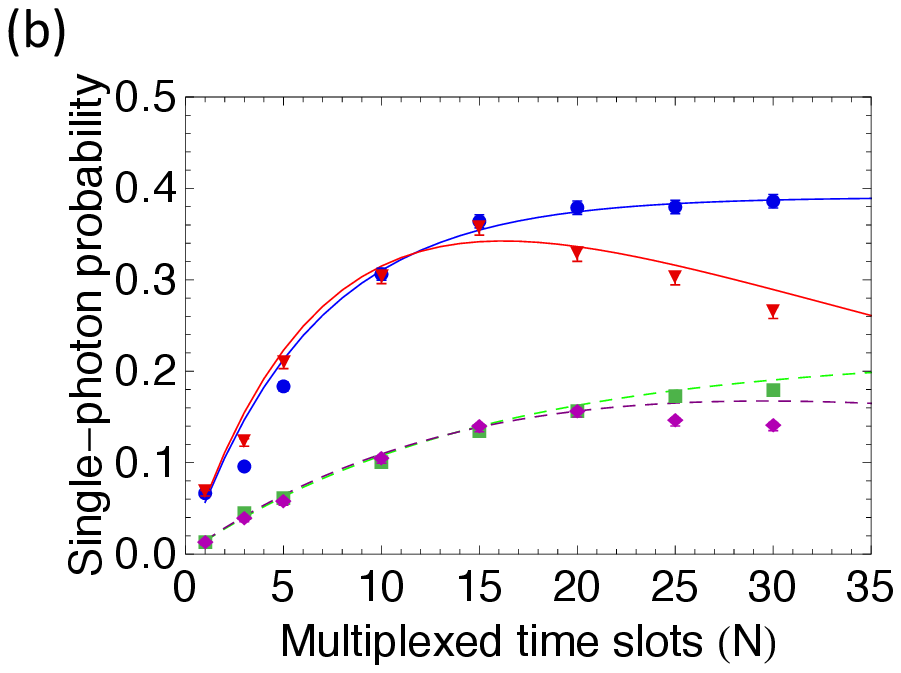}
  \includegraphics[width=0.32\columnwidth,clip]{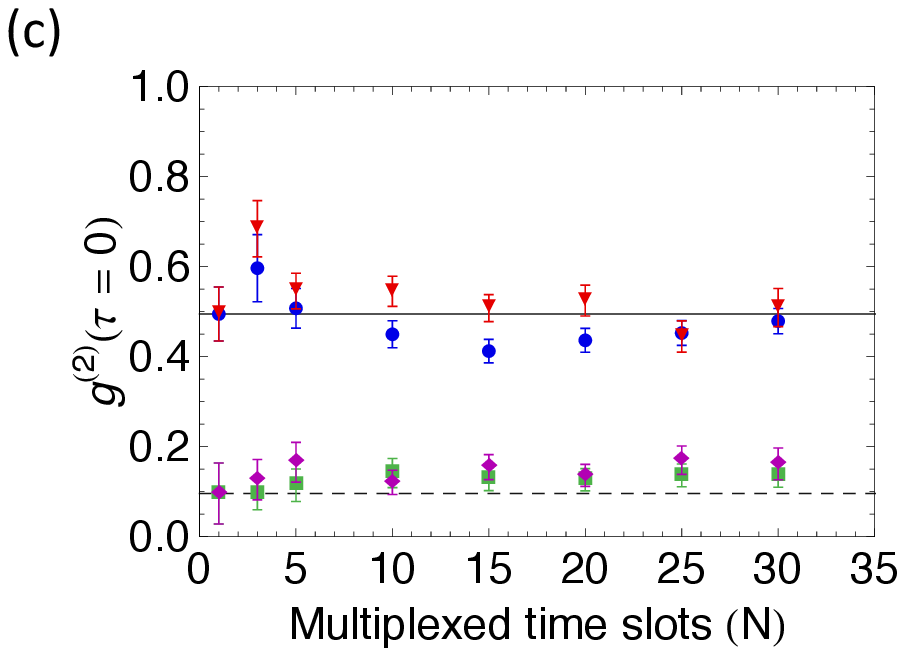}
     \caption{Experimental results. (a) Heralding rate $H$, (b) single-photon probability $P_M(1)$, and (c) the second-order correlation function $g^{(2)}(0)$ versus number of multiplexed time slots $N$. Blue circles: $p = 0.35$, $\Delta t_{DL}\sim$400 ns. Red triangles: $p = 0.35$, $\Delta t_{DL}\sim$200 ns. Green squares: $p = 0.07$, $\Delta t_{DL}\sim$400 ns. Purple diamonds: $p = 0.07$, $\Delta t_{DL}\sim$200 ns. Error bars are estimated by poissonian photon counting statistics. 
For (a) and (b), solid and dashed curves show theoretical predictions for the measurement with $p = 0.35$ and 0.07, taking into account the measured experimental parameters ($p, T_c, T_i, \eta_{Ds}$) and using Eq. (3) in Supplementary Information. Solid and dashed lines in (c) indicate the values of  $g^{(2)}(0)$ measured without multiplexing ($N$ = 1) for $p = 0.35$ and 0.07. 
}
\label{fig.model}
\end{figure}
\end{tiny}

We studied time multiplexing with two different pump powers: A high pump power (such that $p = 0.35$) is used so that the heralding signal rate $H$ is nearly saturated, i.e., $H$ is close to the repetition rate of the multiplexed source $R = 50$ kHz for $N = 30$ (see blue circles and red triangles in Fig. 3 (a)). 
With a low pump power ($p = 0.07$), $H$ increases almost linearly with $N$ up to 30 (see green squares and purple diamonds in Fig. 3 (a)). 
Note that $R$ for this proof-of-principle multiplexed source is limited by the duty cycle of the PC drivers. 
Figure 3 (b) shows our experimental results for single-photon-in-periodic-output-time-slot $P_M(1)$ versus $N$. 
$P_{M}(1)$ is estimated by 
\begin{align}
P_{M}(1) = \frac{S_1+S_2}{R \eta_{Di}}, 
\label{PM1_exp} 
\end{align}
where $S_1$ and $S_2$ are the single-photon count rates measured by SPD1 and SPD2 at the output time windows, and $\eta_{Di}$ is the system detection efficiency for the multiplexed photons (see Methods). 
For all the measurement in our experiment, we clearly observed an enhancement in $P_{M}(1)$ above the single-cycle probability: 
We observed $P_{M}(1) = 38.6 \pm  0.4$\% for $p = 0.35$ and $N = 30$ (see blue dots in Fig. 3 (b)), corresponding $\sim6$ times enhancement over the non-multiplexed heralded photons (for $N = 1$, $P_{M}(1) = 6.8 \pm 0.4$\%). 
The enhancement factor was even larger for $p = 0.07$ ($\sim$16, see green squares in Fig. 3 (b)), because $H$ increases as $N$ up to 30. 
To our knowledge, these single-photon probability and enhancement factors are superior to all previous demonstrations \cite{Ma:2011in,Collins:2013eu} of multiplexed heralded single-photons (that achieved $P_{M}(1) < 1$\% with enhancement factor of 1.6-4). 
For the results with the 200-ns delay line (red triangles and purple diamonds in Fig. 3 (b)), as expected $P_M(1)$ starts to decrease after $N \sim$15, due to increased loss for photons stored in the earlier heralded time slots. 
Our results are in excellent agreement with our theoretical prediction (solid and dashed curves in Fig. 3(b)), except the observed single-photon probabilities with the 200-ns delay line for $N > 20$. 
The difference may be caused by slight spatial-mode mismatching between the output of the delay line and the switchable storage cavity---a 1\% mismatch would be sufficient to make slight beam clipping at the PC in the storage cavity for $N \gtrsim$ 20 and to account for the observed discrepancy.  

We characterized the ratio of single- and multi-photon probabilities via the second-order correlation function at zero time delay between the signals from SPD1 and SPD2 \cite{Loudon2000}: 
\begin{align}
g^{(2)}(0) = \frac{C_{12}H}{S_{1} S_2},  
\label{g2_exp}
\end{align}
where $C_{12}$ is the coincidence count rate between SPD1 and SPD2. 
The observed $g^{(2)}(0) <$ 1 (see Fig. 3 (c)) is expected for a non-classical light source. 
Moreover, while the single-photon probability is greatly enhanced by multiplexing, the observed $g^{(2)}(0)$ is nearly constant versus $N$ for the same $p$: the contribution of the two-photon probability relative to the single-photon one is approximately independent of the time multiplexing. 
As expected (solid and dashed lines in Fig. 3 (c)), $g^{(2)}(0)$ is higher for $p =$ 0.35, because of the higher likelihood of multi-pair emissions compared to the case for $p =$ 0.07. 
Nevertheless, our results are still better than what one could achieve even from an ideal heralded SPDC source (e.g., with 100\%-efficient trigger detection and perfect coupling into lossless optics, see Table I), which predicts $P_{M}(1) = 36.8$\% (25.0\%) for a completely mixed (pure) single-photon state according to poissonian (thermal) photon-number statistics \cite{Christ:2011ku}. 

There are several straightforward improvements for the time-multiplexed source. 
First, we observed a low purity (0.05) of the heralded idler-photon state (using the method proposed in \cite{Christ:2011ku}), because of strong spectral entanglement. 
Since many QIP applications utilize multi-photon interference effects, which require pure states \cite{Hong:1987vi},
our SPDC source must be changed to one that heralds single photons in a pure quantum state, i.e., without spectral entanglement between the signal and idler photons. 
Sources that herald intrinsically pure single photons without narrowband filtering, have been reported \cite{Mosley:2008ki, Evans:2010jn, Spring:2013tv}; combined with temporal multiplexing, these have the potential for both high single-photon probability and purity.
%
(Note that the probability of higher-order emissions can actually be higher for pure heralded single-photon sources than for ones generating mixed states, because of their different photon-number statistics, i.e., thermal for a single-mode SPDC source versus poissonian for multimode states \cite{Christ:2011ku}; multiplexing is consequently even more important to suppress unwanted multiple-photon events). 

Beyond preparing pure photons, it is important to maintain this purity throughout the time-multiplexing process. 
In particular, group velocity dispersion (GVD) in the storage cavity needs to be negligibly small, lest the cycle-dependent GVD disturb the spectral-temporal mode purity. 
Thus, more careful dispersion management in the storage cavity is required for heralded photons with shorter pulse durations ($\lesssim 100$ fs).  
Finally, SPDs with higher detection efficiencies would better detect unwanted multiple photons in the signal mode, improving the likelihood of heralding true single-photon states. 
Recent state-of-the-art detector technologies, such as visible-light photon counters \cite{Kim:2013ul} and arrays of superconducting nanowire detectors \cite{Marsili:2013fs} have achieved $> 90$\% detection efficiencies and timing resolutions below 100 ps. 

With feasible values for collection (95\%) and detector efficiency (90\%), and delay-line (1\%) and switchable cavity losses (1\%), we anticipate single-photon probability up to 80\%, with a $g^{(2)}(0)$ as low as 0.05 (see Table I). 
Despite that our current source rate is only 50 kHz, such performance would already be advantageous for building up larger photon numbers; for example, the generation rate of an 8-photon state would be $0.8^8 \times 50$ kHz$\sim$8000 per second by using the multiplexing technique, while recent 8-photon experiments \cite{Pan:2012kv} with non-multiplexed SPDC sources generated only $\sim$1 per second.

In conclusion, we experimentally demonstrated a time-multiplexed heralded single-photon source. 
Our experiment is implemented by a highly efficient SPDC source, low-loss fixed optical delay line, and adjustable storage cavity with fast polarization switches. 
We observed large enhancement in single-photon probabilities by multiplexing up to 30 time slots of the heralded single-photon source. 
Despite the great enhancement in the single-photon probability, the observed second-order correlation function of the output photons showed that the ratio of single-photon and unwanted two-photon probabilities for the multiplexed source was as low as for a non-multiplexed one. 
We anticipate that incorporating a bright and pure heralded single-photon source and state-of-the-art photon-number-resolving detectors will enable a periodic and near-deterministic single-photon source that will be a critical component for larger scale optical QIP applications.

\begin{center}
\begin{table}[t]
\caption{Comparision between non-multiplexed sources and multiplexed sources. 
The ideal non-multiplexed sources assume photon-pair sources with perfect heralding efficiency and 100\%-efficient detectors with photon-number-resolving functionality. 
For the time-multiplexed source with efficient components, the values are predicted by feasible parameters ($p$ = 0.1, $T_c$ = 99\%, $T_i$ = 95\%, $\eta_{Ds}$ = 90\%) using recent state-of-the-art technologies and Eq. (S1) in Supplementary Information.  }
\begin{tabular}{ l  cc | cc}      
                     &\multicolumn{2}{c}{Ideal non-multiplexed source}            & \multicolumn{2}{|c}{Multiplexed source}    \\
                     & Mixed state                                 & Pure state                  & This work                   & With efficient components   \\  \hline 
  $P_{M}(1)$   & 36.8 \%                                       & 25.0 \%                     & 38.6 $\pm$ 0.4 \%     & $>$ 80\% \\
  $g^{(2)}(0)$ & 0                                                 & 0                               & 0.479  $\pm$ 0.028  & $<$ 0.05      \\ \hline
\end{tabular}

\end{table}
\end{center}


\section*{Methods}
We used a frequency-tripled mode-locked Nd:YAG laser ($\lambda = 355$ nm, $\Delta t \sim$5 ps, $\tau = 8.33$ ns) to pump a bismuth barium borate (BiBO) crystal that generates non-collinear photon pairs centered at 710 nm via Type-I phase-matching. 
Each photon produced by the SPDC was passed through an interference filter (IF, $\Delta \lambda =  20$ nm), and then coupled into a single-mode fiber (SMF), with a coupling efficiency above 70\% \cite{Christensen:2013ux}. 
We used a cascade of four SPDs, each a silicon avalanche photodiode (Si-APD), with detection efficiency of $\sim$70\%. 
The detector cascade allows us to reduce the effect of each SPD's saturation and to resolve the approximate number of photon pairs produced in a time slot. 
By measuring the ratio of single- and coincidence-count rates from the SPDC source, we estimate that the total system detection efficiency of the signal photons $\eta_{Ds} = 41.8 \pm 0.4$\%. 
A field-programmable gate array (FPGA) module processes input signals from the SPDs, and triggers PCs at heralded and output time slots. 
A Herriott-cell-type optical delay line \cite{Robert:2007tj} is made of two concave spherical mirrors separated by $\sim$1 m. 
The idler photon enters the cell through a hole in one mirror, bounces 54 times between the two mirrors, and escapes from the cell through a hole in the other mirror. 
We used one way (round trip) of the Herriott cell to implement $\Delta t_{DL} \sim$200 ns ($\sim$400 ns), with transmission of 89.3\% (84.6\%). 
For an adjustable photon storage cavity, we used a custom Brewster-angled polarizing beamsplitter (PBS1) and PC comprising a pair of rubidium titanyl phosphate (RTP) crystals. 
The single-pass cavity transmission ($T_c = 97.0$\%) is limited by imperfect reflection of mirrors and PBS1, and the PC's imperfect transmission and polarization switching. 
A PC and PBS2 after the cavity are used as an optical shutter, which opens in an output time window of the multiplexing only if at least one heralding signal is detected within $N$ multiplexed time slots, and drops unheralded photons. 
The total transmission of other optics for the idler mode $T_i$ is 92.4\%. 
In order to investigate the multiplexed single-photon probability using Eq. \eqref{PM1_exp}, we estimated the system detection efficiency for the multiplexed photon $\eta_{Di}$ = 56.8\%, accounting for the measured transmission of the non-polarizing 50:50 beam splitter (NPBS, 89.9\%), fiber coupling efficiency and transmission (90.2\%), and SPD detection efficiency ($70\%$, as listed in manufacturer specification sheet).

\section*{Acknowledgement}
This work was supported by NSF Grant No. PHY 12-12439, US Army ARO DURIP grant No. W911NF-12-1-056, and the MURI Center for Photonic Quantum Information Systems (ARO/ARDA Program DAAD19-03-1-0199). 

%

\section*{Supplementary Information}
Here we show theoretical details of a time-multiplexed single-photon source with inefficient optical components. 
As discussed in the main text, for the case that multiple time slots within $N$ are heralded, it is the more efficient to use the photon that is created last.  
The probability of emitting an $m$-photon state by multiplexing $N$ time slots and storing the last-born photons is given by
\begin{align}
 P_M(m) =& \sum_{j =1 }^N ( 1 - \sum_{k_1 =1 }^{\infty}  P_c(k_1)P_d(1|k_1))^{N-j+1} \sum_{k_2 = 1 }^\infty P_c(k_2)P_d(1|k_2) P_e(m|j, k_2), 
\label{Pl(m)}
\end{align}
where $P_c(k)$, $P_d(1|k)$, and $P_e(m|j, k)$ are, respectively, the probability of generating $k$ photon pairs in a time slot, detecting a signal photon conditioned by $k$ photon pairs created, and emitting $m$ idler photons conditioned by $k$ photon pair generations in the $j$-th time slot. 
The first part in Eq. \eqref{Pl(m)} is the probability that a single signal photon detection has not occurred in the ``last'' $j-1$ time slots. 
The second part is the probability that a single signal photon is detected at the $j$-th time slot, and $m$ photons are released at the output time window. 
For pure heralded photons via SPDC generating separable photon pairs, $P_c(k)$ follows a thermal distribution \cite{Christ:2011ku}: $P_c(k) = p^k/(1+p)^{k+1}$, where $p$ is the mean number of created photon pairs per time slot. 
However, the observed purity of our heralded idler photons is Tr$\rho^2 \sim$0.05, indicating that our SPDC source generates photon pairs into $s \sim1/{\rm Tr}\rho^2 \sim$20 Schmidt modes. 
Assuming that each of $s$ Schmidt modes has the identical mean photon numbers per time slot $p/s$, $P_c(k)$ follows the distribution: 
\begin{align}
 P_c(k) = \frac{(p/s)^k}{(1+p/s)^{k+s}}\binom{k+s-1}{k}. 
 \label{Pc_s}
\end{align}
For $s \gg 1$, and thus for our SPDC source, Eq. \eqref{Pc_s} closely follows a Poissonian distribution: 
\begin{align}
 P_c(k) \simeq \frac{e^{-p}p^k}{k!}. 
 \label{Pc_poisson}
\end{align}
The detection probability of a signal photon $P_d(1|k)$ is given by
\begin{align}
 P_d(1|k) &= \sum_{l =1 }^k \eta_{Ds}^l (1-\eta_{Ds})^{k-l} \binom{k}{l} \left(\frac{1}{D}\right)^{l-1}, 
\label{Pd}
\end{align}
where $\eta_{Ds}$ is the total transmission of the signal photons from the SPDC crystal to SPDs, and $D$ is the number of SPDs used for a trigger-detector cascade to herald idler photons. 
Here, we assume that the SPDs are ``bucket'' detectors that discriminate between zero and one-or-more photons, and the detector cascade distributes the signal photons to $D$ detectors with an equal probability ($1/D$). 
By substituting $D \rightarrow  \infty$, Eq. \eqref{Pd} can also describe a photon-number-resolving detector: $ P_d(1|k) =  \eta_{Ds} (1-\eta_{Ds})^{k-1}$. 
The idler photon's transmission to an output time slot $P_e(m|j, k)$ is given by
\begin{align} 
P_e(m|j,k) &= (T_i T_c^{N-j})^m (1-T_i T_c^{N-j})^{k-m} \binom{k}{m}. 
\label{Pe}
\end{align}
The $T_c$ and $T_i$ denote the idler photon's storage efficiency in the switchable storage cavity and other optics transmission (including an initial delay line, fiber coupling efficiency, etc.), respectively. 
For the case without an extra optical delay line to determine the last heralded idler photon, $P_M(m)$ for storing the earliest-born photons is given by \cite{Jeffrey:2004ky}: 
\begin{align}
 P_M(m) =& \sum_{j =1 }^N ( 1 - \sum_{k_1 =1 }^{\infty}  P_c(k_1)P_d(1|k_1))^{j-1} \sum_{k_2 = 1 }^{\infty} P_c(k_2)P_d(1|k_2) P_e(m|j,k_2).  
\label{Pf(m)}
\end{align}
The above equation is similar to Eq. \eqref{Pl(m)} except that the first part in Eq. \eqref{Pf(m)} describes the probability that the single signal photon is not detected in the ``first'' $j-1$ time slots. 
As shown in Fig. 3 (c) in the main text, Eq. \eqref{Pf(m)} has an optimal number of time slots $N_{\rm max}$ to maximize $P_M(m)$, because increasing $N$ effectively results in larger photon storage times, and associated larger loss in the storage cavity. 
$N_{\rm max}$ is obtained by solving $\partial P_M(m)/ \partial N = 0$ in $N$: 
\begin{align} 
N_{\rm max} &\simeq \frac{\log (\log (1- \alpha)) -\log (\log T_c) }{\log T_c -\log (1-\alpha)}, \\
\alpha &= \sum_{k_1 =1 }^{\infty}  P_c(k_1)P_d(1|k_1). 
\label{Nopt}
\end{align}
For our experiment, $N_{\rm max}\sim$14 and $\sim$29 for $p = 0.35$ and 0.07, respectively.

\end{document}